\documentclass{iopart}

\usepackage{graphicx}

\newcommand{\hide}[1]{}

\begin{document}

\title[Neutrons from fragmentation of light nuclei: GEANT4 study]{Neutrons from fragmentation 
of light nuclei in tissue-like media: a study with GEANT4 toolkit}
    
\author{Igor~Pshenichnov$^{1,2}$, Igor~Mishustin$^{1,3}$ and Walter~Greiner$^1$}  
\address{$^1$ Frankfurt Institute for Advanced Studies, Johann Wolfgang Goethe University, 
60438 Frankfurt am Main, Germany}
\address{$^2$ Institute for Nuclear Research, Russian Academy of Science, 117312 Moscow, Russia}
\address{$^3$ Kurchatov Institute, Russian Research Center, 123182 Moscow, Russia}

\begin{abstract}
We study energy deposition by light nuclei in tissue-like media
taking into account nuclear fragmentation reactions, in particular, 
production of secondary neutrons. The calculations are carried out 
within a Monte Carlo model for Heavy-Ion Therapy (MCHIT) based on the GEANT4 toolkit.
Experimental data on depth-dose distributions for 135A-400A MeV 
$^{12}$C and $^{18}$O beams are described very well without any 
adjustment of the model parameters. 
This gives confidence in successful use of the GEANT4 toolkit for MC simulations of
cancer therapy with beams of light nuclei. 
The energy deposition due to secondary neutrons produced by $^{12}$C and $^{20}$Ne
beams in a (40-50 cm)$^3$ water phantom is estimated to 1-2\% of the total dose,
that is only slightly above the neutron contribution ($\sim$1\%) induced by a 200 MeV proton beam.
\end{abstract}

%\submitto{\PMB}
%\pacs{87.53.Pb, 87.53.Wz, 87.53.Vb} 
% Key words: theory and algorithms in physics of heavy-ion therapy, Monte Carlo applications, simulation

\ead{pshenich@fias.uni-frankfurt.de}

\section{Introduction}

Using charged particles from accelerators for cancer treatment was first proposed by
R.R.~Wilson (Wilson 1946)\hide{~\cite{Wilson:1946}}, who noticed that protons and light nuclei have a unique 
property of delivering maximum dose at the end or their range in matter (Bragg peak). 
The centenary of the Bragg's discovery of a sharp peak in the depth-dose curve
for alpha particles in air (Bragg 1905)\hide{~\cite{Bragg:1905}} was celebrated recently, in particular, by 
publishing a dedicated paper (Brown and Suit 2004)\hide{~\cite{Brown:and:Suit:2004}}. 
The precise localization of high doses in 
the region of the Bragg peak suits well the aim of killing cancer cells in  
deeply-sitting localized tumors 
while sparing surrounding healthy tissues. This aim cannot be achieved with the 
conventional radiotherapy using ionizing gamma, electron or neutron irradiations 
which all are characterized by a broad depth-dose distribution without any sharp maximums.
 
Only a few decades after the Wilson's proposal, in the sixties, the advancements in accelerator 
techniques made possible a wide use of the proton beams for therapeutic purposes 
(Archambeau \etal 1974, Orecchia \etal 1998)\hide{~\cite{Archambeau:et:al:1974,Orecchia:et:al:1998}}.
In the seventies alpha-particles 
(Castro \etal 1997, Nowakowski \etal 1991)\hide{~\cite{Castro:et:al:1997,Nowakowski:et:al:1991}} and neon 
ions (Linstadt \etal 1991, Castro \etal 1994)\hide{~\cite{Linstadt:et:al:1991,Castro:et:al:1994}} 
were first used for patient treatments.  
In the last ten years remarkable results were obtained 
in cancer therapy with carbon ions at the  Gesellschaft f\"{u}r Schwerionenforschung
GSI (Schulz-Ertner \etal 2004, Kr\"{a}mer \etal 2003)\hide{~\cite{Schulz-Ertner:et:al:2004,Kraemer:et:al:2003}} 
in Germany, and at the National Institute of 
Radiological Sciences at HIMAC facility (Tsujii \etal 2004)\hide{~\cite{Tsujii:et:al:2004}} in Chiba, Japan. 

As concluded by Tsujii \etal (2004)\hide{~\cite{Tsujii:et:al:2004}}, 
carbon-ion beams provide physical and biological 
advantages over photons. Heavy ions propagating through the tumor produce 
much higher local density of secondary electrons as compared with photon and even proton
irradiation (Kraft 2000, Kraft \etal 1999, Brons \etal 2003)\hide{~\cite{Kraft:2000,Kraft:et:al:1999,Brons:et:al:2003}}. 
This enhances the mortality of cancer cells. 
The accumulated positive experience made possible the creation of the first 
European dedicated medical center for proton, helium- and carbon-ion 
therapy in Heidelberg (J\"{a}kel \etal 2003)\hide{~\cite{Jakel:et:al:2003}}. 

After the advantages of heavy ions in killing tumor cells are clearly demonstrated, 
the next important step is to work out the methods for 
precise delivery of high doses to tumors while minimizing irradiation of 
normal tissues. Developing efficient treatment procedures requires joint efforts of physicists, biologists,  
medical doctors, accelerator engineers and computer experts.  It is obvious that a common computational tool   
is needed for exchanging information and accumulating experience obtained by different 
hadron and heavy-ion therapy centers.

Up to now mostly deterministic methods were used by different 
groups for dose calculations in hadron therapy treatment planning, see e.g. 
Hong \etal (1996), Kr\"{a}mer \etal (2000), 
J\"{a}kel \etal (2001)\hide{~\cite{Hong:et:al:1996,Kramer:et:al:2000,Jakel:et:al:2001}}. 
They use empirical information about various processes, for instance 
fragmentation cross sections 
(Schardt \etal 1996, Sihver \etal 1998)\hide{~\cite{Schardt:et:al:1996,Sihver:et:al:1998}}, 
to calculate the depth-dose distributions for specific treatment conditions.
Therefore, these deterministic methods can not be easily extended for calculating the 
doses in irradiation conditions different from those where the experimental data were taken and fitted.
For example, the one-dimensional computer codes developed for proton therapy can not be immediately used for
heavy-ion treatment planning.  The important advantages of deterministic methods consist in 
their relative simplicity and low computational time.

Alternatively, several three-dimensional Monte Carlo approaches are available now which can be used 
for particle transport calculations in highly inhomogeneous human tissues with air caverns and bones.
The lateral scattering of beam particles as well as production and transport of secondary particles 
can be easily taken into account in such calculations. The Monte Carlo approach was implemented 
in PETRA (Medin and Andreo 1997)\hide{~\cite{Medin:and:Andreo:1997}} and 
PTRAN (Berger 1993)\hide{~\cite{Berger:1993}} codes, which are dedicated to the simulation of
proton transport in tissue-like media, with beam energies of interest for cancer therapy. 
Several general-purpose Monte Carlo transport codes were 
tested for dose calculations from proton beams 
(Fippel and Soukup 2004, Paganetti and Gottschalk 2003, 
Paganetti 2004, 
Jiang and Paganetti 2004)\hide{~\cite{Fippel:2004,Paganetti:and:Gottschalk:2003,Paganetti:2004,Jiang:and:Paganetti:2004}}. 
In these evaluation studies VMCpro (Fippel and Soukup 2004)\hide{~\cite{Fippel:2004}}, 
FLUKA (Fass\'{o} \etal 2000)\hide{~\cite{Fasso:et:al:2000}}, 
GEANT3 (Goosens and Giani 1994)\hide{~\cite{Goosens:and:Giani:1994}} 
or GEANT4 (Agostinelli \etal 2002)\hide{~\cite{Agostinelli:et:al:2002}} 
codes were used. In particular, the role of nuclear interactions of primary 
and secondary particles in proton therapy was studied by 
Paganetti (2002)\hide{~\cite{Paganetti:2002}} by using the 
GEANT3 code. 

A detailed review of the transport codes capable to describe the 
propagation and fragmentation of heavy ions in extended media was given 
recently by Gudowska \etal (2004)\hide{~\cite{Gudowska:et:al:2004}}. The results of the SHIELD-HIT 
code (Gudowska \etal 2004)\hide{~\cite{Gudowska:et:al:2004}} 
were presented in details. The authors conclude that their code describes well 
most of the presently available experimental data and thus can be used for Monte Carlo 
simulations of heavy-ion therapy. Another particle and heavy ion transport code PHITS 
was modified recently (Nose \etal 2005)\hide{~\cite{Nose:et:al:2005}} to
improve calculations of dose distributions in carbon therapy. Unfortunately, both of the 
two codes are not publicly available. In particular, as explained by 
Gudowska \etal (2004)\hide{~\cite{Gudowska:et:al:2004}}, 
SHIELD-HIT code is currently under further development and it is not available to the general user. 

To the best of our knowledge, no evaluation studies on heavy-ion therapy have been done
with Monte Carlo models which are based on modern publicly available software tools.
It seems very promising for this purpose to use a recent version 7.0 of the GEANT4 
toolkit (Agostinelli \etal 2002)\hide{~\cite{Agostinelli:et:al:2002}}. This toolkit is widely used in experimental 
high-energy and nuclear physics for simulating complicated particle 
detectors (GEANT4-Webpage 2005)\hide{~\cite{GEANT4-Webpage:2005}}.
Recently this toolkit was updated by including realistic models for nuclear interactions
at intermediate energies of interest for heavy-ion therapy. 
In this paper we apply it for simulating transport of protons, carbon, oxygen and neon 
ions in tissue-like media. This approach provides a self-consistent 
way for calculating the production rates of secondary nuclear fragments, 
protons and neutrons as well as electrons and photons. 

First, the total depth-dose distributions are calculated and compared with
experimental depth-dose curves. In contrast to most previous studies, here they are 
calculated with correct absolute (not relative) normalization. 

Second, we study the energy deposition due to nuclear reactions induced by secondary nucleons and fragments,
with a focus on the secondary neutrons. Our study makes possible the direct comparison of
neutron contributions to the total energy deposition for proton and heavy-ion beams.

\section{Physical models in GEANT4}

In the present study we used the version 7.0 (with patch 01) of the GEANT4 
toolkit (GEANT4-Webpage 2005)\hide{~\cite{GEANT4-Webpage:2005}} 
to build a Monte Carlo model for Heavy-Ion Therapy (MCHIT).
In this model a tissue-like phantom, which is irradiated by proton or 
heavy-ion beam, is presented by a water cube. We stress that in this paper we study the 
secondary particle production only in the water phantom, and leave aside
possible nuclear reactions in the beam control elements which are usually located 
in front of the phantom.

A detailed description of physical models included in GEANT4 is given in 
the Physics Reference Manual (GEANT4-Documents 2005)\hide{~\cite{GEANT4-Documents:2005}}. 
The set of models which are most relevant to a particular simulation 
problem should be activated by a user.
In order to make this selection, one can start either from the  
available examples of previously developed applications, or from the so-called predefined physics lists.
In the both cases specific physical processes, the low-energy neutron capture,  for example, 
can be activated or de-activated in a simulation via corresponding
user interface commands. The examples and predefined physics lists are provided by 
GEANT4 developers (GEANT4-Webpage 2005)\hide{~\cite{GEANT4-Webpage:2005}} 
and distributed along with the GEANT4 source code. 
 
For MCHIT we choose the set of models called ``standard electromagnetic physics'' 
to describe the energy losses of primary and secondary charged particles 
due to electromagnetic processes. The models of ionization energy loss and 
multiple scattering were activated for all charged particles. The simulations 
are based on the Bethe-Bloch formula taking into account the density effect and 
shell corrections (GEANT4-Documents 2005)\hide{~\cite{GEANT4-Documents:2005}}.

In each simulation step the ionization energy loss of a charged particle is calculated according to
the Bethe-Bloch formula. If this energy loss is higher than a threshold for the emission 
of $\delta$-electrons (which is taken as an input parameter in the
calculation) the emission of a $\delta$-electron is simulated. 
In the opposite case the energy loss is subtracted from the kinetic energy of the particle and
then added to the calculated local energy deposition.

One of the novel features of GEANT4 consists 
in the possibility for user to specify the production thresholds in therms of secondary particle ranges 
in specified materials. A relatively high cut in 
range of 1 mm  for secondary electrons, positrons and photons was used in simulations,
that corresponds to the lowest energies 350 keV, 340 keV and 3 keV of these particles, respectively.
At the same time the maximum step size in track calculations was taken as small as 0.1 mm.
The continuous local energy deposition was calculated for each of these small steps thus 
providing a very detailed depth-dose distribution. The selected options have noticeably 
reduced the CPU time without affecting the accuracy of calculation of the depth-dose
distribution.  
  
The bremsstrahlung processes for electrons and positrons were also activated in the calculations along with
the annihilation process for positrons. For photons, the Compton scattering, the conversion into an  
electron-positron pair and the photoelectric effect were considered.    

In the present evaluation study  the option of low-energy electromagnetic 
physics available in GEANT4 (GEANT4-Webpage 2005)\hide{~\cite{GEANT4-Webpage:2005}} 
was not used, although it can be easily utilized in the future. 
By using the low-energy models,
one can reduce the production threshold for $\delta$-electrons down to 200 eV.
The low energy processes include the photoelectric effect, Compton scattering, Rayleigh scattering, 
gamma conversion, bremsstrahlung and ionization. The fluorescence of excited atoms is also considered. 
The low-energy models make direct use of the atomic-shell cross section data, 
while the standard calculations rely on parameterizations of these data.

GEANT4 simulations include two kinds of hadronic interactions:
(a) elastic scattering of hadrons on target protons and nuclei, which dominate 
at low projectile energies, and (b) inelastic nuclear reactions induced by 
fast hadrons and nuclei (Agostinelli \etal 2002)\hide{~\cite{Agostinelli:et:al:2002}}.    
The model of nucleon-nucleon elastic scattering is based on parameterization of experimental data in
the energy range of 10-1200 MeV. The Monte Carlo simulation of elastic scattering is quite simple, since
the final states of these reactions are identical with the initial ones. 
At higher energies the hadron-nucleus elastic
scattering can be simulated within the Glauber model (GEANT4-Documents 2005)\hide{~\cite{GEANT4-Documents:2005}}.  

The modeling of inelastic interactions of hadrons and 
nuclei requires more serious efforts. In the GEANT4 toolkit one is able to make his own choice of 
models to be involved in simulations. Two groups of models are available for inelastic interactions of hadrons:  
(a) the data driven models, which are based on the parameterizations of measured 
cross sections for specific reaction channels,   
and (b) the theory driven models, which are based on various theoretical approaches and implemented
as Monte Carlo event generators (Agostinelli \etal 2002)\hide{~\cite{Agostinelli:et:al:2002}}. 

In the MCHIT model the inelastic interaction of low-energy (below 20 MeV) nucleons, including radiative
neutron capture, were simulated by means of data driven models. Above 20 MeV the exciton-based precompound
model was invoked (Agostinelli \etal 2002)\hide{~\cite{Agostinelli:et:al:2002}}. 
Finally, an intranuclear cascade model -
the binary cascade model (Folger \etal 2004)\hide{~\cite{Folger:et:al:2004}} was used for 
energies above 80 MeV for fast hadrons and nuclei.
Since exited nuclear remnants are created after the first cascade stage of interaction, appropriate
models for describing the de-excitation process have to be involved into simulation.   
The Weisskopf-Ewing model (Weisskopf and Ewing 1940)\hide{~\cite{Weisskopf:and:Ewing:1940}} 
was used for the description of 
evaporation of nucleons from excited nuclei at relatively low excitation energies, 
while the Statistical Multifragmentation Model (SMM) (Bondorf \etal 1995)\hide{~\cite{Bondorf:et:al:1995}}  was used
at excitation energies above 3 MeV per nucleon
to describe multi-fragment break-up of highly-excited residual nuclei.
The SMM model includes as its part the Fermi break-up model, which describes the explosive decay of
highly-excited light nuclei.

\section{Linear energy deposition of ions in water}

\subsection{Total depth-dose distribution}

\begin{figure}[htb]  
\begin{centering}
\includegraphics[width=0.8\columnwidth]{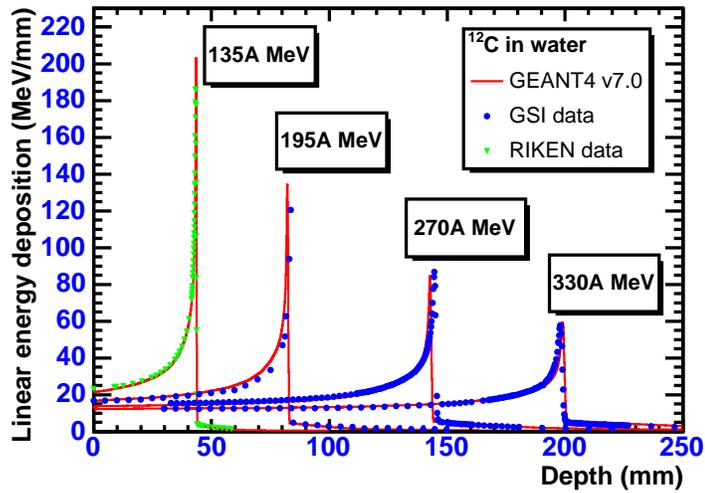}
\end{centering}
\caption{\label{fig:1} Average linear energy deposition by $^{12}$C ions in water.
The beam energies are given in the boxes. GEANT4 calculations are shown by histograms,
experimental data from GSI (Sihver \etal 1998)\hide{~\cite{Sihver:et:al:1998}} 
and RIKEN (Kanai \etal 1993)\hide{~\cite{Kanai:et:al:1993}} 
are shown by circles and triangles, respectively.}
\end{figure}

\begin{figure}[htb]  
\begin{centering}
\includegraphics[width=0.8\columnwidth]{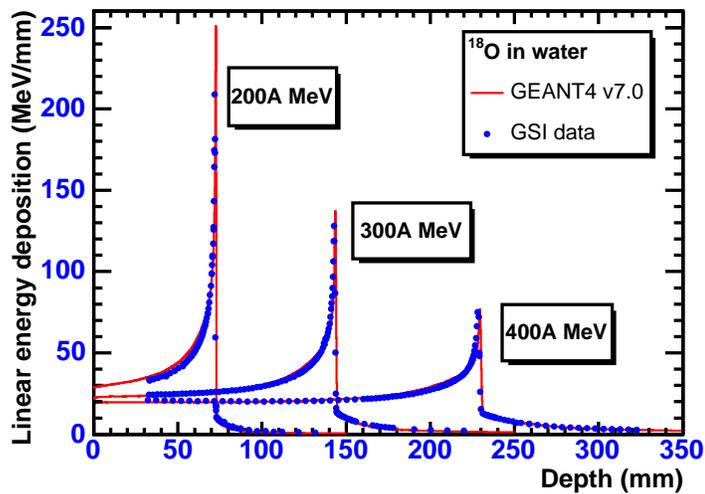}
\end{centering}
\caption{\label{fig:2} Same as in figure~\ref{fig:1} but for $^{18}$O ions.}
\end{figure}

\begin{figure}[htb]  
\begin{centering}
\includegraphics[width=0.9\columnwidth]{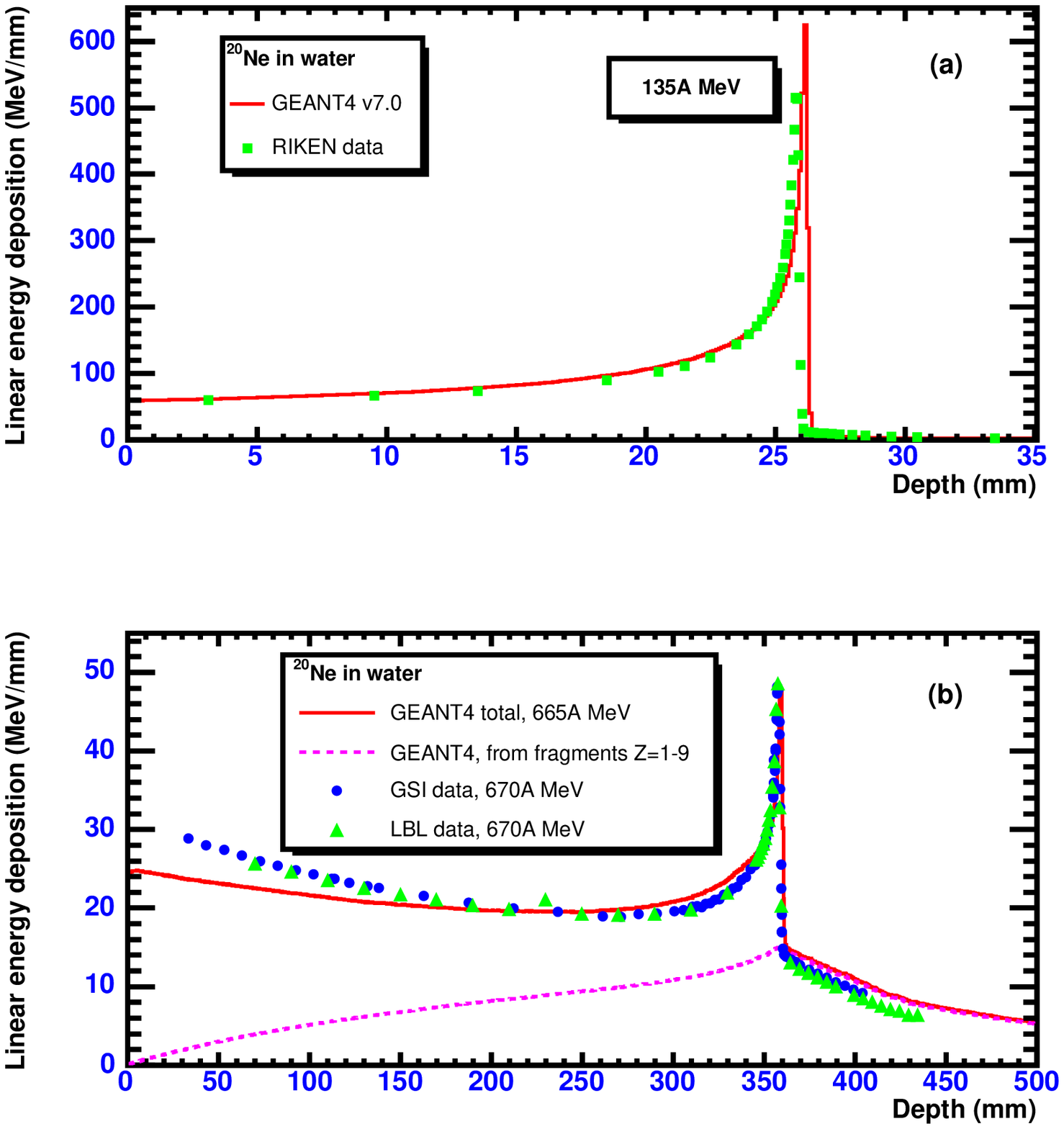}
\end{centering}
\caption{\label{fig:3} Same as in figure~\ref{fig:1} but for 135A MeV (a) and 670A MeV (b) $^{20}$Ne ions.
RIKEN (Kanai \etal 1993)\hide{~\cite{Kanai:et:al:1993}}, 
GSI (Sihver \etal 1998)\hide{~\cite{Sihver:et:al:1998}} and 
LBL (Wilson \etal 1984)\hide{~\cite{Wilson:et:al:1984}} data are shown by 
squares, circles and triangles, respectively.}
\end{figure}

The depth-dose distribution from proton and ion beams propagating through 
the tissue-like media is the most important characteristic to be considered for 
cancer therapy applications. As discussed above, the beams of 
light nuclei (carbon or oxygen) are very promising for this purpose. In order to quantify 
the amount of energy deposited by an ion along its track in matter, one usually considers the average 
linear energy deposition $\rmd E/\rmd x$ or the average linear energy transfer (LET), expressed in 
MeV/mm, or keV/${\mu}$m. This characteristics was calculated by splitting a cubic 
phantom into thin slices and calculating the
energy deposited in each of the slices. The resulting depth-dose curves for $^{12}$C, $^{18}$O and 
$^{20}$Ne ions in water are presented in figures~\ref{fig:1},\ref{fig:2} and \ref{fig:3}.
In the plots the calculated values are given in MeV/mm, 
while the experimental data, which are usually not available in 
the absolute units, are rescaled to be superimposed on the theoretical curves.

It is a well-known fact that the charged particle range in a material of a given chemical composition 
is inversely proportional to the density of electrons, and hence, to the matter density $\rho$.     
It is quite common to present the depth-dose distributions as functions of the areal density $x\cdot\rho$,
since $\rmd E/\rmd x$ at other density can be easily obtained by rescaling these distributions.
The depth-dose curves presented in figures~\ref{fig:1}, \ref{fig:2} and \ref{fig:3} 
were calculated for pure water phantoms with the water density of 
$\rho=1.0$ g/cm$^3$ at 4$^0$ C. Therefore, the obtained distributions numerically have the same 
scale in depth $x$, as the distributions given in the areal density scale $x\cdot\rho$ 
expressed in g/cm$^2$. It was supposed that experimental distributions were obtained 
for pure water at 20$^0$ C ($\rho=0.998203$ g/cm$^2$),
and experimental data were appropriately rescaled.


One can conclude that the present approach based on the GEANT4 toolkit 
provides a very good description of the depth-dose curves in water for $^{12}$C and $^{18}$O ions. 
The shapes of distributions are well
reproduced, while the peak position is reproduced with the accuracy of 1-2 mm.  
One of the main calculational parameters in the Bethe-Bloch formula, which controls the exact 
position of the Bragg peak, is the average ionization potential $I$ for the water molecule. 
In our simulations we fix it to $I=70.89$ eV, a default value used in GEANT4, which is calculated by
the appropriate values for hydrogen and oxygen atoms.     

As seen in figure~\ref{fig:3}a, the experimental depth-dose curves for $^{20}$Ne ions 
are satisfactorily described only at 135A MeV,
but a noticeable discrepancy between theory and data appears at 670A MeV in figure~\ref{fig:3}b.
Keeping in mind possible uncertainties in the
incoming beam energy, one can try to fine-tune the beam energy in calculations. 
We have found that the peak position is better reproduced with the beam energy of 665A MeV.  
This is in line with the results reported by Gudowska \etal (2004)\hide{~\cite{Gudowska:et:al:2004}}, 
where the authors used the SHIELD-HIT code to describe the same data and found that 
the experimental data are better described with the beam energy of 667A MeV. 
 
One can note that 670A MeV $^{20}$Ne ions are less suitable for heavy ion therapy due to increased
fragmentation. Indeed, the peak value is only two times higher than the entrance dose. 
This will result in higher doses in healthy tissues on the way to deeply-seated tumors.

\subsection{Fragmentation reactions}

\begin{figure}[htb]  
\begin{centering}
\includegraphics[width=0.9\columnwidth]{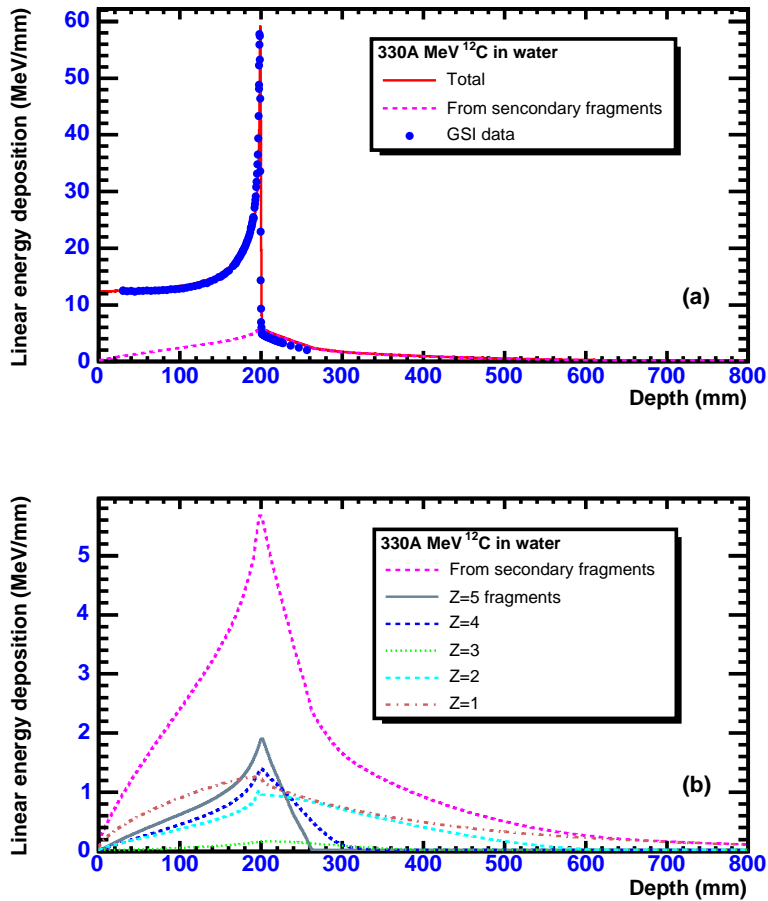}
\end{centering}
\caption{\label{fig:4} Contribution to the total energy deposition (a) from secondary 
fragments (b) produced in nuclear interactions of 
 330A MeV carbon ions in water. The experimental data from GSI (Sihver \etal 1998)\hide{~\cite{Sihver:et:al:1998}} 
are shown by circles.}
\end{figure}

In case of heavy-ion beams it is important to study the processes 
responsible for the energy deposition beyond the Bragg peak, which are entirely due to the
nuclear fragmentation reactions. In calculations presented in figure~\ref{fig:4} the average linear energy
deposition was determined separately for the fragments of each charge. 
Our analysis shows that due to their longer
ranges, protons and alpha particles are mostly responsible for the doses beyond the Bragg peak.
The doses from heavier fragments with $Z=4,5$ are mostly localized around the stopping point of the initial
$Z=6$ nuclei.

The fact that the energy deposition due to secondary fragments is peaked at exactly the Bragg peak
of primary ions (see figure~\ref{fig:4})  has a simple explanation. 
Initially, when primary nuclei are still energetic enough to induce 
fragmentation reactions, the flux of secondary fragments grows and accordingly, 
their energy deposition grows too. However, beyond the range of primary $^{12}$C ions the 
production of secondary fragments stops. Therefore, the dose curves for secondary fragments fall down 
after that point due to the decreasing number of such fragments.

\section{Energy deposition from secondary neutrons}

Neutrons as neutral particles do not lose their energy directly via ionization, but rather via 
the secondary reactions involving charged hadrons.
They can be produced either in neutron elastic scattering on target nuclei or via
inelastic nuclear reactions. The mean free path of fast neutrons in matter is typically much longer 
as compared to the ranges of protons of the same energy, since the protons suffer a direct energy
loss due to ionization of target atoms, but the neutrons do not.
Therefore, the dose associated with neutrons is expected to be distributed in a much larger volume 
compared to protons. In particular, this property of neutron beams was exploited in neutron therapy 
experiments (Orecchia \etal 1998)\hide{~\cite{Orecchia:et:al:1998}}.
     
Since the secondary neutrons can be produced in both proton and heavy-ion irradiations, it is
important to compare their effects in these two cases. 
The role of secondary neutrons produced in the interaction of protons with thick targets 
was investigated previously in several experimental and theoretical works. 
As example, we mention here some recent studies aimed at 
medical applications (Schneider \etal 2002, Wroe \etal 2005)\hide{~\cite{Schneider:et:al:2002,Wroe:et:al:2005}}. 

Less is known about the doses from the secondary neutrons created
in heavy-ion irradiations. In this case one can expect that the fast neutrons created in
fragmentation of projectile nuclei, the so-called spectator neutrons, provide an 
additional dose as compared with the proton irradiations. 
In two recent experiments 
(Gunzert-Marx \etal 2004, Kumamoto \etal 2005)\hide{~\cite{Gunzert-Marx:et:al:2004,Kumamoto:et:al:2005}}
neutron detectors were placed behind the targets and the effective doses and attenuation lengths in
shielding materials were measured along with the energy and angular distributions of neutrons.
The obtained information is indispensable for designing the neutron shielding for 
heavy-ion therapy rooms.

However, to the best of our knowledge, there are no publications reporting the doses from  
secondary neutrons {\em inside} the extended target or phantom in heavy-ion irradiations.
Since a measured depth-dose distribution gives the total dose, it includes
the local energy deposition from charged hadrons and nuclear fragments from neutron-induced 
reactions. Due to the fact that secondary neutrons are expected 
to go far beyond the Bragg peak, the dose from neutrons has to be calculated in a wider region, 
i.e. outside the normal region of depth-dose measurements.

\subsection{Neutron induced reactions}

In contrast to fast protons and nuclear fragments, which continuously transfer their
energy to $\delta$-electrons while propagating through the medium, 
neutrons lose their energy in a more discrete way.
At some points of their tracks neutrons make elastic or inelastic hadronic interactions
resulting in fast secondary protons or nuclear fragments which thus produce ionization.   
This multistep mechanism of energy transfer from neutrons to electrons explains in general the
biological effects of neutron radiation.

\begin{figure}[htb]  
\begin{centering}
\includegraphics[width=0.8\columnwidth]{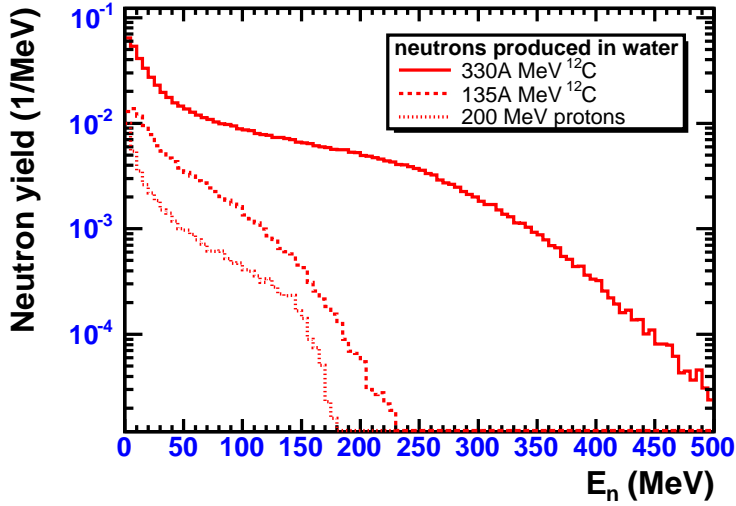}
\end{centering}
\caption{\label{fig:neutrons} Kinetic energy spectra of secondary neutrons 
produced by 135A and 330A MeV carbon ions and 200 MeV protons in a (40 cm)$^{3}$ water cube.
The spectra are normalized to one beam particle and contain all neutrons 
produced at different points of the ion track and at different angles.}
\end{figure}

In figure~\ref{fig:neutrons} we present kinetic energy spectra of secondary neutrons 
calculated with the GEANT4 toolkit. As seen in the figure, the spectrum of secondary neutrons
is rather wide, since several neutron production mechanisms are involved. It contains
low-energy neutrons emitted by excited target nuclei, high-energy neutrons from
the projectile evaporation as well as intermediate energy neutrons from the 
overlapping parts of nuclei (fireball).

Let us first consider the neutrons produced from the target nuclei.
As a rule, the fragments of target nuclei have low velocity in the laboratory system,
but they can be highly excited. Low-energy neutrons with kinetic energies of a few MeV, typically below 10 MeV,
are produced in the de-exitation process of the remnants of target nuclei via preequlibrium 
emission or evaporation. For light nuclei like $^{16}$O, the same mechanism is in effect 
also for protons.

\begin{figure}[htb]  
\begin{centering}
\includegraphics[width=0.8\columnwidth]{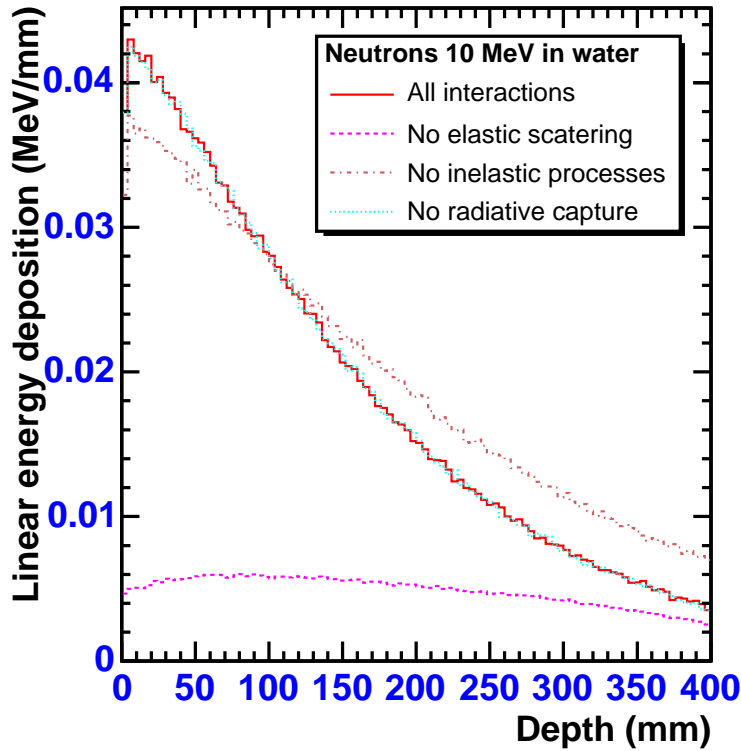}
\end{centering}
\caption{\label{fig:5} Average linear energy deposition by 10 MeV neutron beam in (40 cm)$^3$ water phantom.
The full calculation with all the neutron interaction channels enabled is presented 
by the solid histogram. The calculations where all $(n,n')$ channels, 
inelastic neutron-induced reactions or the radiative neutron capture process are disabled
are shown by dashed, dashed-dotted and dotted histograms, respectively.}
\end{figure}

Elastic scattering (including charge-exchange reactions) on target protons and 
nuclei ($^{16}$O) are the main processes responsible for the energy loss by 
low-energy neutrons in water. In oder to 
estimate the relative contribution of elastic and inelastic hadronic interactions to the 
energy loss of neutrons, one can calculate the energy deposition from a {\em monoenergetic} neutron beam in
a rather thick (40 cm)$^3$ water phantom. The results of our GEANT4 calculations are given in 
figure~\ref{fig:5} for 10 MeV neutrons. In this case the absolute values of average linear energy deposition
are below 0.4 MeV/mm, which is less than $\sim$0.5\% of the total energy deposition from ions
before the Bragg peak, see figure~\ref{fig:4}a. Nevertheless, it is at the level of 
a few percent compared to the average linear energy deposition 
{\em after} the Bragg peak, see figure~\ref{fig:4}b. 

As shown in figure~\ref{fig:5}, in the case when elastic reactions are disabled,
the energy deposition from 10 MeV neutrons is essentially reduced compared to the full
calculation. On the contrary, the contribution from the radiative neutron capture 
reactions $^{16}{\rm O}(n,k\gamma)^{17}{\rm O}$ with $k=1,2,3...$, as predicted by the 
GEANT4 simulation, is negligible for the 10 MeV neutron beam. When the inelastic nuclear reactions,
$n+^{16}{\rm O}\rightarrow n+^{16}{\rm O}+\gamma$,
$n+^{16}{\rm O}\rightarrow n+^{13}{\rm C}+\alpha+\gamma$, 
$n+^{16}{\rm O}\rightarrow 4\alpha+n$, ... are disabled,
the attenuation of the initial neutron flux is somehow reduced, leading to a less steep fall of
the depth-dose curve. However, the average energy deposition is still comparable to the case of
full calculation.

\begin{figure}[htb]  
\begin{centering}
\includegraphics[width=0.9\columnwidth]{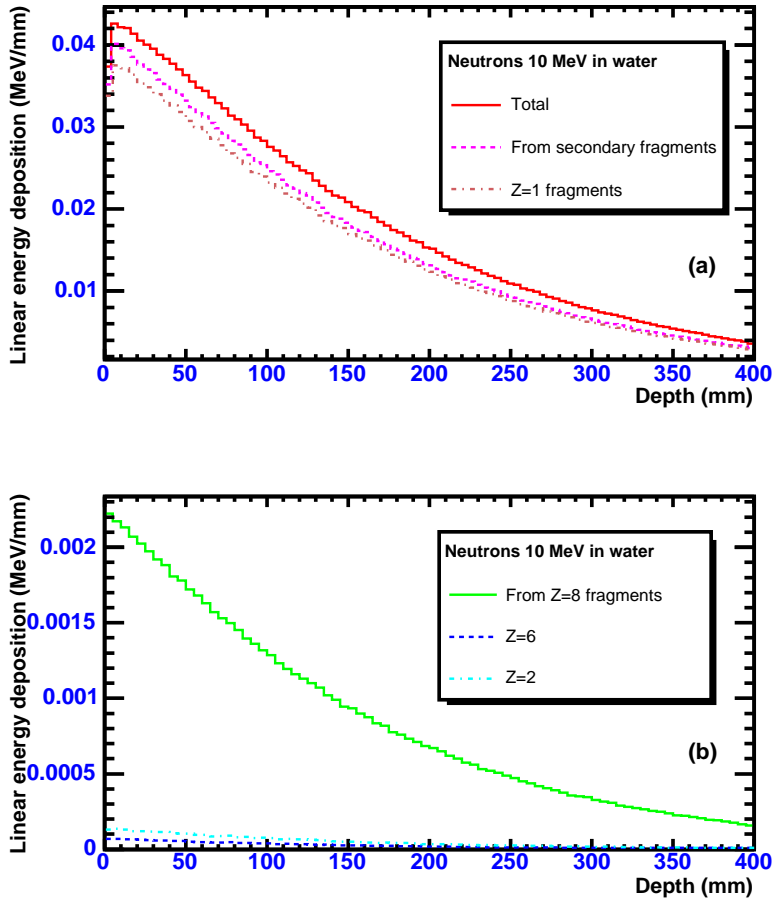}
\end{centering}
\caption{\label{fig:5frag} Contributions from secondary fragments to the average linear energy deposition 
by 10 MeV neutrons in (40 cm)$^3$ water phantom.
The total energy deposition (solid histogram), the contribution from secondary 
fragments with $Z=1-8$ (dashed histogram) and from protons (dash-dotted histogram) are shown in 
panel (a). The contributions from $Z=8$ (solid histogram), $Z=6$ (dashed histogram) and $Z=2$ (dash-dotted
histogram) fragments are shown in panel (b).}
\end{figure}

The role of secondary protons and nuclear fragments in the energy dissipation of
10 MeV neutrons is demonstrated in figure~\ref{fig:5frag}. The distributions were obtained by
calculating directly the energy deposition from secondary protons and nuclear fragments 
separately for each particle charge. 
The sum of such contributions is nearly identical to the total energy deposition with
a small difference originating from the contribution of fast secondary electrons. As seen in
figure~\ref{fig:5frag}a, secondary $Z=1$ fragments (mostly protons) are responsible for
80-90\% of the energy deposition of 10 MeV neutrons. The contribution from recoiling $^{16}$O nuclei
from elastic neutron scattering and the contribution from $^{12}$C and $^{4}$He fragments created
in neutron-induced $^{16}$O dissociation reactions are much smaller. They are shown in
figure~\ref{fig:5frag}b.

As seen in figure~\ref{fig:5}, the average linear energy deposition of 10 MeV neutrons 
is decreasing with increasing depth in water. 
Low-energy neutrons transfer their kinetic energy to recoil protons and nuclei. 
These slow charged particles have ranges of few millimeters in water, 
and the calculated energy deposition simply reflects the attenuation of the flux and energy of
primary neutrons propagating through the water.

\begin{figure}[htb]  
\begin{centering}
\includegraphics[width=0.8\columnwidth]{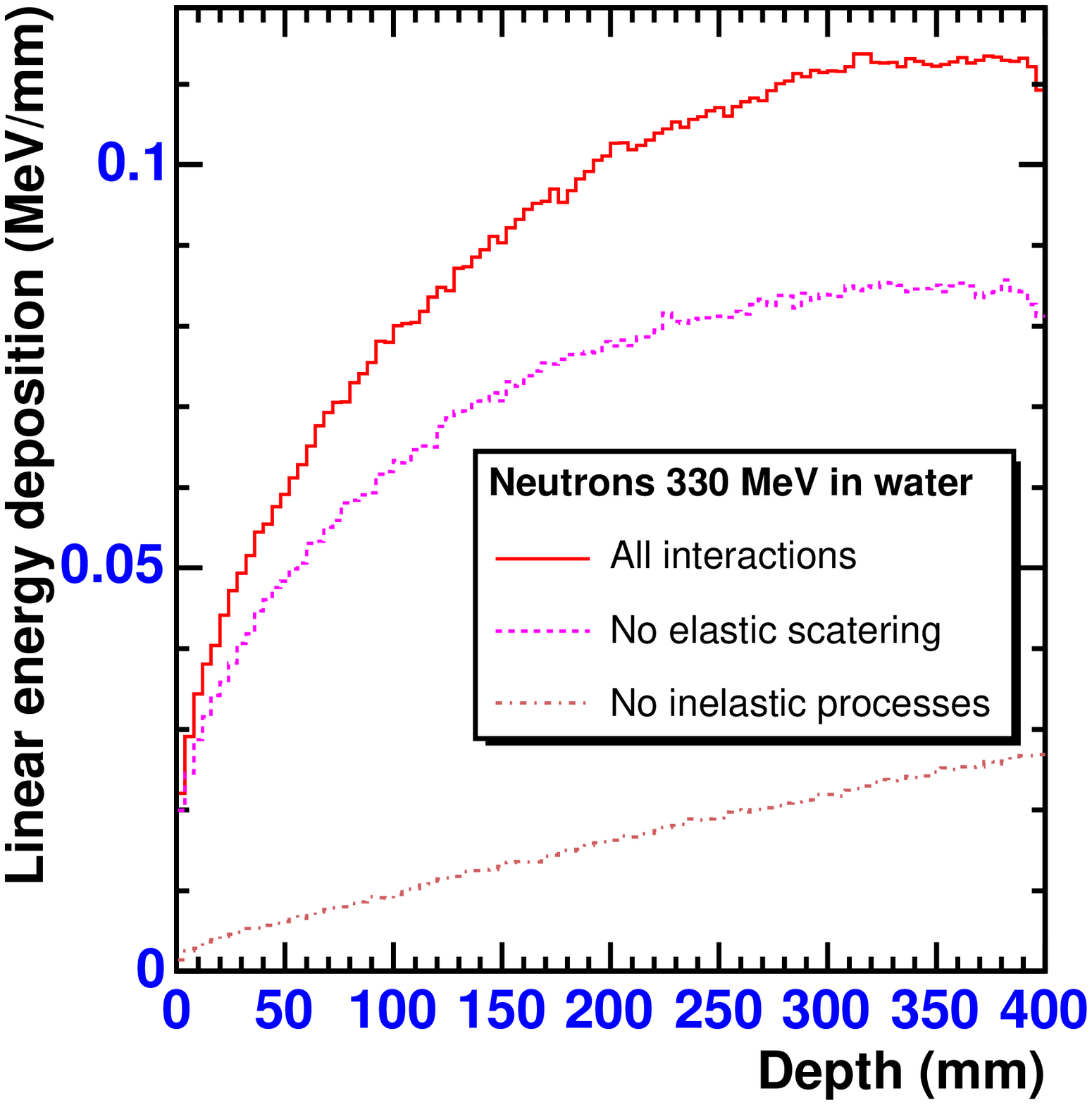}
\end{centering}
\caption{\label{fig:6} Average linear energy deposition by 330 MeV neutron beam in (40 cm)$^3$ water phantom.
The full calculation with all neutron interaction channels enabled is presented 
by the solid histogram. The calculations where all $(n,n')$ channels, or 
inelastic neutron-induced reactions are disabled shown by dashed and dashed-dotted
histograms, respectively}
\end{figure}

Fast neutrons, with energies close to the beam energy per nucleon, originate from the projectile 
fragmentation reactions. The mechanism of their energy deposition in water is different from one for 
low-energy neutrons, as demonstrated in figure~\ref{fig:6} for 330 MeV neutrons.
The average linear energy deposition is increasing with the depth in phantom, because 
the secondary protons and nuclear fragments have long ranges in water, comparable to the
phantom size, and they stop far from the production point.
In other words, a fast neutron create a kind of particle shower, which needs some space for its
development. In addition, the cross sections for neutron interaction with protons and nuclei
are generally lower at 330 MeV compared to 10 MeV neutrons. Therefore, the attenuation of the neutron flux
is less pronounced at distances up to 40 cm shown in the figure~\ref{fig:6}.    
 
\begin{figure}[htb]  
\begin{centering}
\includegraphics[width=0.9\columnwidth]{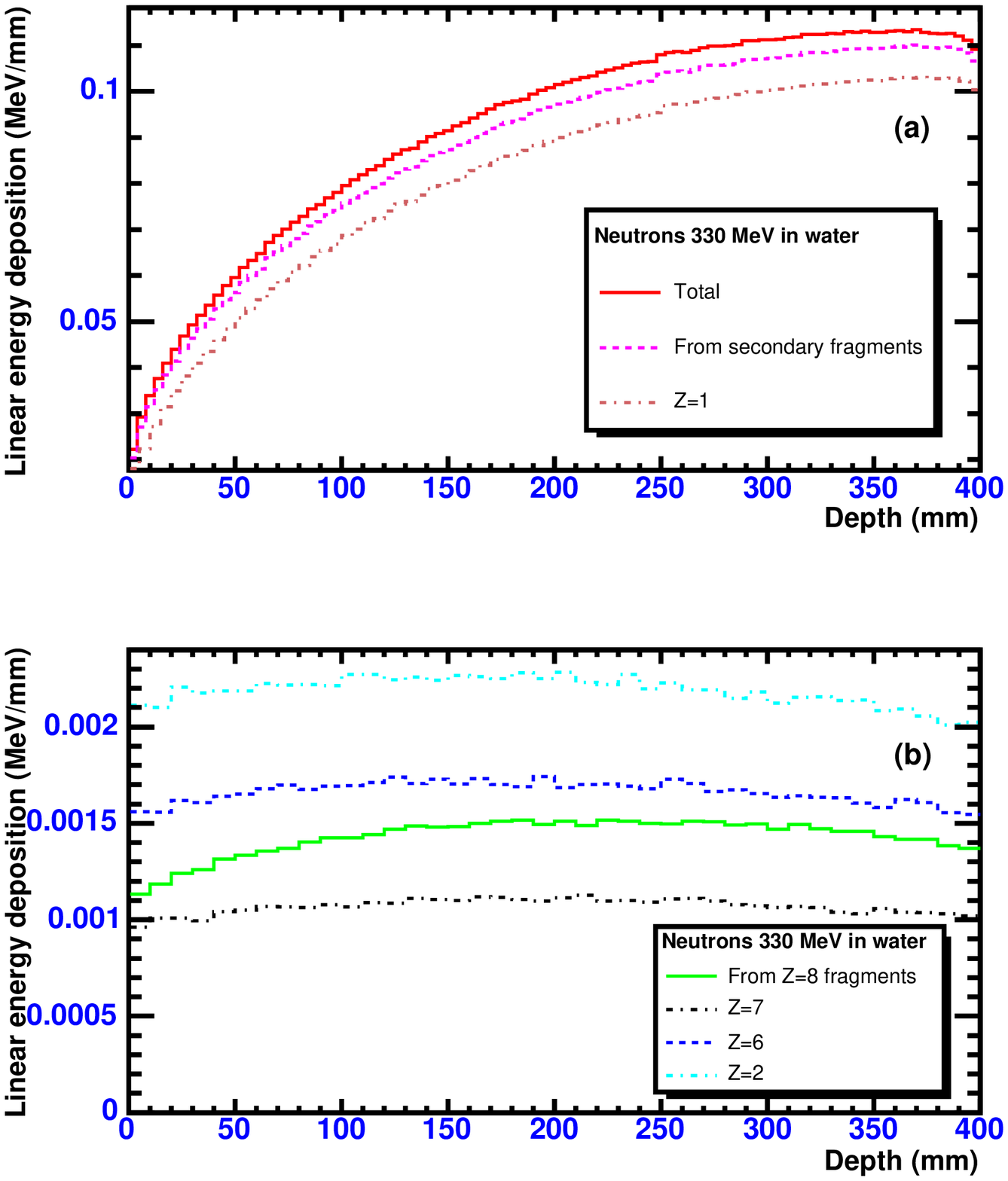}
\end{centering}
\caption{\label{fig:6frag} Same as in figure~\ref{fig:5frag}, but for 330 MeV neutrons. 
In addition, the contribution from secondary fragments with $Z=7$ is shown by the thick dash-dotted
line in panel (b).}
\end{figure}

When the elastic scattering on target protons and oxygen nuclei is neglected,
the energy deposition of 330 MeV neutrons 
is only slightly reduced compared to the full calculation. However,
when inelastic neutron-induced reactions are neglected, the energy deposition is reduced dramatically. 
Therefore, the neutron-induced inelastic reactions with oxygen nuclei are responsible for 
the main part of energy deposition of fast spectator neutrons.      

The contributions to the dose from secondary protons and nuclear fragments 
for 330 MeV neutron beam are shown in figure~\ref{fig:6frag}.
Again, the distributions were obtained by calculating directly the energy deposition 
from secondary protons and nuclear fragments separately for each particle charge. 

As seen in figure~\ref{fig:6frag}a, the $Z=1$ fragments (mostly protons) are also 
responsible for 80-90\% of the energy deposition of fast neutrons, but 
in this case such protons are produced mainly in inelastic neutron-induced nuclear reactions.
Several protons can be produced in a single neutron-induced reaction on $^{16}$O. 
The transfer of neutron energy to more heavy 
fragments of oxygen nuclei with $Z=2-8$ is less efficient and leads to much lower energy deposition via
these fragments compared to protons, see figure~\ref{fig:6frag}b.   
The absolute values of energy deposition
from 330 MeV neutrons are approximately twice as large compared with 10 MeV neutrons, 
but still less than $\sim$1\% of  the energy deposition from primary ions
before the Bragg peak, see figure~\ref{fig:4}a for comparison.

As follows from figures~\ref{fig:5} and~\ref{fig:6}, despite the energy depositions of low- and 
high-energy neutrons are comparable, the spatial distributions of corresponding doses are very different.
The low-energy secondary neutrons produced by proton or heavy-ion beams deposit their
energy in the vicinity of a primary track. On the contrary, secondary high-energy neutrons deposit energy in 
a large volume, which is comparable to the phantom size, but their number is low.

\subsection{Comparison of neutron-induced doses in proton and heavy-ion irradiations}

After the mechanisms of energy dissipation of monoenergetic neutrons in water are clarified, 
one can directly compare the role of secondary neutrons in proton and heavy-ion irradiations. 
We consider three typical irradiation setups: (I) 200 MeV protons in
(40 cm)$^3$ water cube, (II) 330A MeV $^{12}$C ions in (40 cm)$^3$ water cube and 
(III) 670A MeV $^{20}$Ne ions in (50 cm)$^3$ water cube. These cases reflect 
typical distances in human body, since the ranges of beam particles 
are $\sim26$ cm, $\sim20$ cm and $\sim37$ cm, respectively. 
Therefore, $25$\%-50\% of water thickness is beyond the Bragg peaks that represents schematically
healthy tissues where the irradiation is undesirable.

Three sets of calculations have been performed: (1) calculations with electromagnetic interactions
only, i.e. without hadronic interactions and fragmentation reactions,
(2) full calculations with hadronic interactions activated for all hadrons, and
(3) calculations with hadronic interactions activated for all hadrons except neutrons.
In the last case secondary neutrons are allowed to leave the phantom freely without 
interactions.   

The calculation results for the fraction of beam energy deposited in phantoms 
are presented in table~\ref{NeutIntRates}. Since the secondary nucleons and nuclei are not produced 
in the case (1), the beam energy is completely deposited in the phantoms. 
This first case can be also considered as a test demonstrating the accuracy 
of the methods used.

\begin{table}[htb]
\caption{\label{NeutIntRates} 
GEANT4 results for cubic water phantoms irradiated by 200 MeV protons, 330A MeV carbon and 670A MeV neon ions. 
The fraction of beam energy deposited in phantoms (in \%\%) is given 
(1) for calculations with only electromagnetic processes included, (2) for calculations with all processes included, 
(3) for calculations with all hadronic and fragmentation processes included, 
but without interactions of secondary neutrons. 
The fraction of beam energy deposited due to the interactions of secondary neutrons is given 
as the difference between rows (2) and (3).
The last row shows the neutron doses divided by the total doses deposited in phantom and by the 
average RBE of proton (1.1) and heavy-ion (3.0) beams.}
\begin{indented}
\item[]\begin{tabular}{@{}llll}
\br
  &            I              &         II            &         III            \\
  & protons at 200 MeV        & $^{12}$C at 330A MeV  & $^{20}$Ne at 670A MeV  \\
  &  in (40 cm)$^3$ water     &  in (40 cm)$^3$ water &  in (50 cm)$^3$ water  \\
  &    cube                   &        cube           &      cube              \\
\mr
(1) Only            &  &  & \\
electromagnetic &  100. & 100. & 99.997 \\
interactions    &  &  & \\
& & & \\
(2) All processes & & & \\ 
including &   94.48 &  88.63  & 68.25  \\
fragmentation & & & \\
& & & \\
(3) Without & & & \\
neutron &  93.49 &  87.51 & 66.80 \\
interactions & & & \\
& & & \\
Contribution & & & \\
from secondary &  0.99 &  1.12 & 1.45 \\
neutrons & & & \\
& & & \\
{\bf Neutron} & & & \\
{\bf dose divided} &   {\bf 0.95} &   {\bf 0.42} &  {\bf 0.71} \\
{\bf by the total } & & & \\
{\bf dose and RBE} & & & \\

\br
\end{tabular}
\end{indented}
\end{table}

Full calculations, case (2), show that some part of the beam energy is taken away by
the secondary particles produced in hadronic interactions and, in the case of ion beams, in
projectile fragmentation reactions. In the case (III) of 670A MeV $^{20}$Ne ions
more than 30\% of the beam energy is taken away by
secondary particles, mostly protons, as they have
longer ranges in water compared to nuclear fragments. 

As shown in table~\ref{NeutIntRates} for the case (3), the omission of neutron interactions in our MCHIT
calculations leads to only a slight decrease of the energy deposited in the phantoms.  
One can estimate the energy deposition due to secondary neutrons as the difference between
the cases (2) and (3). For incident protons, carbon and neon ions the contributions
from secondary neutrons are 0.99\%, 1.12\% and 1.45\% of the beam energy, respectively.
The same characteristic for 200A MeV carbon 
beam irradiating a 12.78 cm water phantom was indirectly 
estimated by Gunzert-Marx \etal (2004)\hide{~\cite{Gunzert-Marx:et:al:2004}} at the
level below 1\%, which is in good agreement with the present GEANT4 results.

As demonstrated by Kraft (2000)\hide{~\cite{Kraft:2000}} and 
Kr\"{a}mer \etal (2003)\hide{~\cite{Kraemer:et:al:2003}},  
the relative biological effectiveness (RBE) of the
carbon ions is in the range of 2-5 compared 
to 1.0-1.1 for protons (Paganetti \etal 2002a)\hide{~\cite{Paganetti:et:al:2002a}}.
This means, in particular, that the physical dose delivered by the $^{12}$C beam to a patient
should be accordingly smaller, as well as the physical dose from the secondary neutrons. 
This point is illustrated in the last row of table~\ref{NeutIntRates}, where the neutron contributions
are given relative to the total doses in phantom and after the division by the corresponding RBE.

Comparison with other codes can be made for the case of proton beams only.
The results of PETRA (Medin and Andreo 1997)\hide{~\cite{Medin:and:Andreo:1997}}, 
PTRAN (Berger 1993)\hide{~\cite{Berger:1993}} and SHIELD-HIT 
codes for water phantom irradiated by 200 MeV protons were presented 
by Gudowska \etal (2004)\hide{~\cite{Gudowska:et:al:2004}}.  
The integrated energy losses under the depth-dose distributions are 
189.6 MeV (PETRA), 188.9 MeV (PTRAN) and
188.3 MeV (SHIELD-HIT). Our result for 200 MeV protons (case (I)) obtained with MCHIT  is 188.96 MeV, 
that is in very good agreement with the predictions 
of other codes (Gudowska \etal 2004)\hide{~\cite{Gudowska:et:al:2004}}.

\section{Conclusion}

Due to prominent advances in computer technology, there exists 
increasing interest in using Monte Carlo methods for heavy-ion therapy simulations. 
The option of using the GEANT4 toolkit (version 7.0) for this purpose  
was investigated in the present evaluation study. We have composed a new GEANT4 application, MCHIT,
which includes consistently all the major processes relevant to heavy-ion 
propagation in tissue-like media. We have demonstrated that the set of experimental 
data on the depth-dose distributions of
$^{12}$C, $^{18}$O and $^{20}$Ne ions in water 
is generally well described by the MCHIT model based on the GEANT4 toolkit .
This is very promising in the view of possible applications of Monte Carlo methods 
for the treatment planning in heavy-ion cancer therapy. 

The present study was focused on the energy deposition from nuclear 
reactions induced by secondary neutrons. 
It is demonstrated that neutron-induced ionization processes due to secondary
protons and nuclear fragments contribute not more than 1-2 \% of the total energy deposited 
in a (40-50 cm)$^3$ water phantom, depending on the beam energy, the charge and mass of beam ions 
and the phantom size. A similar contribution was estimated for a proton beam irradiation.

It is important, however, that due to the higher RBE of ion beams, the physical dose from
secondary neutrons should be smaller than for proton beams. 
Our main conclusion is that additional neutrons produced by  
heavy-ion beams compared to proton beams   
do not appear to be a serious hazard to human body. 
     
The distributions of positron emitting nuclei, radial distributions of energy deposition and 
detailed analysis of ion fragmentation reactions will be presented in subsequent papers. 
Also, we are planning to estimate the doses from secondary neutrons produced on
beam control elements and in highly inhomogeneous human tissues.

\ack
This work was supported by Siemens Medical Solutions.
We are grateful to Hermann Requardt for stimulating discussions which initiated the present study. 
The discussions with Gerhard Kraft, Thomas Haberer, Michael Scholz,
Dieter Schardt and Horst St\"{o}cker are gratefully acknowledged. Special thanks are expressed to Dieter Schardt and
Oksana Filipenko for providing us the tables of experimental data on depth-dose distributions.

% File for Harvard system

\References

%\bibitem{Agostinelli:et:al:2002} 
\item[] Agostinelli S \etal  (GEANT4 Collaboration) 2003 GEANT4: A simulation toolkit {\it Nucl. Instrum. Meth. A } {\bf 506} 250-303

%\bibitem{Archambeau:et:all:1974 
\item[] Archambeau J O, Bennett G W, Levine G S, Cowen R and Akanuma A  1974 Proton radiation therapy {\it Radiology} {\bf 110} 445-57

%\bibitem{Berger:1993}  
\item[] Berger M J 1993 {\it Proton Monte Carlo Transport Program PTRAN.} {\it National Institute of Standrts and Technology Report} NISTIR-5113

%\bibitem{Bondorf:et:al:1995} 
\item[] Bondorf J P, Botvina A S, Iljinov A S, Mishustin I N and Sneppen K  1995 Statistical multifragmentation of nuclei {\it Phys. Rept.} {\bf 257} 133-221

%\bibitem{Bragg:1905} 
\item[] Bragg W H and Kleemann R  1905 On the alpha particles of radium, and their loss of range in passing through various atoms and molecules {\it Phil. Mag. S.6 } {\bf 10} 318-40

\item[] %\bibitem{Brons:et:al:2003} 
Brons S, Taucher-Scholz  G, Scholz M and Kraft G  2003  A track structure model for simulation of strand breaks in plasmid DNA after heavy ion irradiation {\it Radiat. Environ. Biophys.} {\bf 42} 63-72

%\bibitem{Brown:and:Suit:2004} 
\item[] Brown A and Suit H  2004 The centenary of the discovery of the Bragg peak {\it Radiother. Oncol.} {\bf 73}  265-8

%\bibitem{Castro:et:al:1994} 
\item[] Castro J R, Linstadt  D E, Bahary  J P, Petti  P L, Daftari I, Collier J M, Gutin P H, Gauger G and Phillips T~L  1994  Experience in charged-particle irradiation of tumors of the skull base - 1977-1992 {\it Int. J. Radiation Oncology Biol. Phys.} {\bf 29} 647-55

%\bibitem{Castro:et:al:1997} 
\item[] Castro J R, Char D H, Petti P L, Daftari I K, Quivey J M, Singh R P, Blakely E A and Phillips T L  1997 15 years experience with helium ion radiotherapy for uveal melanoma {\it Int. J. Radiation Oncology Biol. Phys.} {\bf 39} 989-96

%\bibitem{Fasso:et:al:2000} 
\item[] Fass\'{o}  A, Ferrari A,  Ranft J and Sala P R  2000 FLUKA: Status and Prospective for Hadronic Applications {\it In Proc. of the MonteCarlo 2000 Conference, Lisbon, October 23-26 2000, Kling A, Barao F, Nakagawa M, Tavora L, Vaz P - eds.} (Berlin: Springer-Verlag) p~955-60 

%\bibitem{Fippel:2004} 
\item[] Fippel M and Soukup M  2004 A Monte Carlo dose calculation algorithm for proton therapy {\it Med. Phys.} {\bf 31} 2263-73 

%\bibitem{Folger:et:al:2004} 
\item[] Folger  G, Ivanchenko V N and Wellisch J P  2004 The Binary Cascade - nucleon-nuclear reactions {\it Eur. Phys. J. A} {\bf 21} 407-17

%\bibitem{GEANT4-Documents:2005} 
\item[] GEANT4-Documents 2005 {\it http://geant4.web.cern.ch/geant4/G4UsersDocuments/Overview/html/}

%\bibitem{GEANT4-Webpage:2005} 
\item[] GEANT4-Webpage 2005 {\it http://geant4.web.cern.ch/geant4/}

%\bibitem{Goosenss:and:Giani:1994} 
\item[] Goosens M and Giani S 1994 {\it GEANT: Detector desription and simulation tool. CERN Program Library Long Writeup W5013} (Geneva: CERN)

%\bibitem{Gudowska:et:al:2004} 
\item[] Gudowska I, Sobolevsky N, Andreo P, Belkic D and Brahme A  2004 Ion beam transport in tissue-like media using the Monte Carlo code SHIELD-HIT {\it Phys. Med. Biol.} {\bf 49} 1933-58

%\bibitem{Gunzert-Marx:et:al:2004} 
\item[] Gunzert-Marx K, Schardt D and Simon R S  2004 Fast neutrons produced by nuclear fragmentation in treatment irradiations with $^{12}$C beam {\it Radiat. Prot. Dosimetry} {\bf 110} 595-600 

%\bibitem{Hong:et:al:1996} 
\item[] Hong L, Goitein M, Bucciolini M, Comiskey R, Gottschalk B, Rosenthal S, Serago C and Urie M  1996 A pencil beam algorithm for proton dose calculations {\it Phys. Med. Biol.} {\bf 41} 1305-30

%\bibitem{Jakel:et:al:2001} 
\item[] J\"{a}kel O, Kr\"{a}mer M, Karger C P and Debus J  2001 Treatment planning for heavy ion radiotherapy: clinical implementation and application {\it Phys. Med. Biol.} {\bf 46} 1101-16

%\bibitem{Jakel:et:al:2003} 
\item[] J\"{a}kel O, Schulz-Ertner D, Karger C P, Nikoghosyan A and Debus J  2003 Heavy ion therapy: status and perspectives {\it Technology in Cancer Research and Treatment} {\bf 2} 377-87

%\bibitem{Jiang:and:Paganetti:2004} 
\item[] Jiang H and Paganetti H  2004 Adaptation of GEANT4 to Monte Carlo dose calculations based onCT data {\it Med. Phys.} {\bf 31} 2811-18

%\bibitem{Kanai:et:al:1993} 
\item[] Kanai T, Minohara S, Kohno T  \etal  1993 Irradiation of 135 MeV/u carbon and neon beams for studies of radiation biology NIRS-M-92 (HIMAC-004), NIRS, Chiba, Japan, 

%\bibitem{Kraft:2000} 
\item[] Kraft G  2000 Tumor therapy with heavy charged particles {\it Prog. Part. Nucl. Phys.} {\bf 45} S473-544

%\bibitem{Kraft:et:al:1999} 
\item[] Kraft G, Scholz M and Bechthold U  1999 Tumor therapy and track structure {\it Radiat. Environ. Biophys.} {\bf 38} 229-37

%\bibitem{Kramer:et:al:2000} 
\item[] Kr\"{a}mer M, J\"{a}kel O, Haberer T, Kraft G, Schardt D and Weber U  2000 Treatment planning for heavy-ion radiotherapy: physical beam model and dose optimization {\it Phys. Med. Biol.} {\bf 45} 3299-317

%\bibitem{Kraemer:et:al:2003} 
\item[] Kr\"{a}mer M, Weyrather W K and Scholz M  2003  The increased biological effectiveness of heavy charged particles: from radiobiology to treatment planning {\it Technology in Cancer Research and Treatment} {\bf 2} 427-36 

%\bibitem{Kumamoto:et:al:2005} 
\item[] Kumamoto Y, Noda Y, Sato Y, Kanai T and Murakami T  2005 Measurements of neutron effective doses and attenuation lengths for shielding materials at the heavy-ion medical accelerator in Chiba {\it Health Phys. }{\bf 88} 469-79

%\bibitem{Linstadt:et:al:1991} 
\item[] Linstadt D E, Castro  J R and Phillips T L  1991 Neon ion radiotherapy - results of the phase-I/II clinical-trial {\it Int. J. Radiation Oncology Biol. Phys.} {\bf 20} 761-9

%\bibitem{Medin:and:Andreo:1997} 
\item[] Medin  J and Andreo P  1997 Monte Carlo calculated stopping-power ratios, water/air, for clinical proton dosimetry (50-250 MeV) {\it Phys. Med. Biol.} {\bf 42}  89-105

%\bibitem{Nose:et:al:2005} 
\item[] Nose H, Niita K, Hara M, Uematsu K, Azuma O, Miyauchi Y, Komori M and Kanai T  2005 Improvement of three-dimensional Monte Carlo code PHITS for heavy ion therapy {\it J. Nucl. Sci. and Tech.} {\bf 42} 250-55 

%\bibitem{Nowakowski:et:al:1991} 
\item[] Nowakowski V A, Ivery G, Castro J R, Char  D H, Linstadt  D E, Ahn D, Phillips T L, Quivey J M, Decker M, Petti P L and Collier J M  1991 Uveal melanoma - development of metastases after helium ion irradiation {\it Radiology} {\bf 178} 277-80.

%\bibitem{Orecchia:et:al:1998} 
\item[] Orecchia R, Zurlo A, Loasses A, Krengli M, Tosi G, Zurrida S, Zucali P and Veronesi  U  1998 Particle beam therapy (hadrontherapy): basis for interest and clinical experience {\it Eur. J. Cancer} {\bf 34} 459-68

%\bibitem{Paganetti:2002} 
\item[] Paganetti  H  2002 Nuclear interactions in proton therapy: dose and relative biological effect distributions originating from primary and secondary particles {\it Phys. Med. Biol.} {\bf 47} 747-64

%\bibitem{Paganetti:et:al:2002a} 
\item[] Paganetti H, Niemierko A, Ancukiewicz M, Gerweck L E, Goitein M, Loeffler J S, Suit H D  2002a Relative biological effectiveness (RBE) values for proton beam therapy  {\it Int. J. Radiat. Oncol. Biol. Phys.} {\bf 53} 407-21.

%\bibitem{Paganetti:2004} 
\item[] Paganetti H  2004 Four-dimensional Monte Carlo simulation of time-dependent geometries {\it Phys. Med. Biol.} {\bf 21}  N75-81

%\bibitem{Paganetti:and:Gottschalk:2003} 
\item[] Paganetti H  and Gottschalk B  2003 Test of GEANT3 and GEANT4 nuclear models for 160 MeV protons stopping in $CH_2$ {\it Med. Phys.} {\bf 30}  1926-31

%\bibitem{Schardt:et:al:1996} 
\item[] Schardt D, Schall I, Geissel H, Irnich H, Kraft G, Magel A, Mohar M F, M\"{u}nzenberg G, Nickel F, Scheidenberger C, Schwab W and Sihver L  1996 Nuclear fragmentation of high-energy heavy-ion beams in water {\it Adv. Space Res.} {\bf 17} 87-94

%\bibitem{Schneider:et:al:2002} 
\item[] Schneider U, Agosteo S, Pedroni E and Besserer J  2002 Secondary neutron dose during proton therapy using spot scanning {\it Int. J. Radiat. Oncol. Biol. Phys.} {\bf 53} 244-51

%\bibitem{Schulz-Ertner:et:al:2004} 
\item[] Schulz-Ertner D, Nikoghosyan  A, Thilmann  C, Haberer  Th, J\"{a}kel O, Karger C, Kraft G, Wannenmacher  M and Debus J  2004 Results of carbon ion radiotherapy in 152 patients {\it Int. J. Rad. Oncol. Biol. Phys.} {\bf 58} 631-40

%\bibitem{Sihver:et:al:1998} 
\item[] Sihver L, Schardt D and Kanai T  1998 Depth-dose distributions of high-energy carbon, oxygen and neon beams in water {\it Jpn. J. Med. Phys.} {\bf 18} 1-21

%\bibitem{Tsujii:et:al:2004} 
\item[] Tsujii H, Mizoe J, Kamada  T, Baba M, Kato S, Kato H, Tsuji H, Yamada S, Yasuda S, Ohno T, Yanagi T, Hasegawa A, Sugawara T, Ezawa H, Kandatsu S, Yoshikawa K, Kishimoto R and Miyamoto T  2004 Overview of clinical experiences on carbon ion radiotherapy at MRS {\it Radiother. Oncol.} {\bf 73}  S41-9

%\bibitem{Weisskopf:and:Ewing:1940} 
\item[] Weisskopf V E and Ewing D H  1940  On the yield of nuclear reactions with heavy elements {\it Phys. Rev.} {\bf 57} 472-85

%\bibitem{Wilson:1946} 
\item[] Wilson R R 1946 Radiological use of fast protons {\it Radiology} {\bf 47} 487-91 

%\bibitem{Wilson:et:al:1984} 
\item[] Wilson J W, Townsend L W, Bidasaria H B, Schimmerling W, Wong M and Howard J  1984 Neon-20 depth-dose relations in water {\it Health Phys.} {\bf 46} 1101-11

%\bibitem{Wroe:et:al:2005} 
\item[] Wroe A J, Cornelius I M and Rosenfeld A B  2005 The role of nonelastic reactions in absorbed dose distributions from therapeutic proton beams in different medium {\it Med. Phys.} {\bf 32} 37-41

\endrefs

\end{document}